\documentclass[reprint,aps,prb,showpacs]{revtex4-1}
\usepackage[intlimits]{amsmath}
\usepackage{amstext}
\usepackage{amsfonts}
\usepackage{amssymb}
\usepackage{bbm}
\usepackage{graphicx}
\usepackage{color}
\usepackage[colorlinks,citecolor=darkblue,filecolor=darkblue,linkcolor=darkblue,urlcolor=darkblue,%
pdfpagelabels,pdfstartview = FitH,bookmarksopen = true,bookmarksnumbered = true,plainpages = false,hypertexnames = false,breaklinks=true]{hyperref}
\usepackage{hyperref}

\definecolor{darkblue}{RGB}{50,90,140}
\newcommand{\sgn}{\operatorname{sgn}}
\def\nn{\boldsymbol{n}}

\begin{document}
\title{Exact treatment of  planar two-electron quantum dots: effects of anharmonicity on the complexity}
\author{Sebastian ~Schr\"oter$^1$}
\email[Email address: ]{sebastian.schroeter@mytum.de}
\author{Paul-Antoine ~Hervieux$^2$, Giovanni ~Manfredi$^2$, Johannes ~Eiglsperger$^{3}$}
\altaffiliation[Present address: ]{numares GROUP, Josef-Engert-Stra\ss e 9, 93053 Regensburg}
\author{Javier ~Madro\~nero$^{1,4}$}
\affiliation{$^1$Physik Department, TU M\"unchen, 85747 Garching\\
$^2$Institut de Physique et Chimie des Mat\'eriaux, CNRS and
Universit\'e de Strasbourg, BP 43, F-67034 Strasbourg, France\\
$^3$Institut f\"ur Theoretische Physik, Universit\"at Regensburg, D-93040 Regensburg\\
$^4$Faculty of Physics, University of Duisburg-Essen, Lotharstra\ss e 1-21, 47048 Duisburg, Germany}
\begin{abstract}
  Static properties of an anharmonic potential model for planar two-electron quantum dots are investigated using a method which allows for the exact representation of the matrix elements, including the full Coulombic electron - electron interaction. The anharmonic confining potential in combination with the interparticle Coulomb interaction affects the spectral properties of the system considerably as it implies total loss of separability of the system. Properties of the classical phase space, spectral measures of the chaoticity, as well as localization properties of the eigenstates corroborate this.
\end{abstract}
\date{\today}
\maketitle

\section{Introduction}
The understanding of the complexity of a physical system has been addressed for the last decades and its physical relevance is still a matter of investigation. For open systems the correlation of quantum chaos and transport is widely discussed, e.g. for quantum dots \cite{Kat04}, giving also evidence for the onset of decoherence and the quantum to classical transition \cite{Fer11}. The connection between thermalization and classically chaotic dynamics is well known since the investigations of ergodicity in the Sinai billiard \cite{Sin70}, while the link between quantum thermalization and quantum chaos is  not clear yet \cite{Rig08}. Experimental techniques for the coupling of microcavities to waveguides exploit the underlying chaoticity of the system \cite{Son12}.

One of the simplest realizations of a complex system in atomic physics is the three-body Coulomb problem. The electron-electron interaction term in the Hamiltonian of
the unperturbed helium atom renders the two-electron
dynamics in general chaotic with only rather small regions of
the classical phase space occupied by regular motion. On the
quantum level, the loss of integrability leads to an abundance of intriguing
and surprising effects \cite{Tan00,Jia08,Eig09a}.
Replacing the attractive Coulomb potential by an alternative kind of potential generates different, though in many cases equally challenging physical models.

The Hooke's atom -- that is two electrons with harmonic confinement -- has been thoroughly investigated, not only because it serves as the most common model for a generic semiconductor quantum dot. Its relative motion can be separated from the center of mass (COM) motion, a fact generally known as Kohn's theorem \cite{Koh61}. Nevertheless, the classical dynamics of the three-dimensional system is in general mixed regular-chaotic \cite{Naz02}. On the quantum level, there is a reduced family of eigenstates, the energies of which admit closed analytical solutions for special values of the confining harmonic potential, the Coulomb interaction between the particles and possibly an additional magnetic field perpendicular to the plane \cite{Tau94a}. 

Due to its simplicity, the harmonic two-electron quantum dot has been used as a paradigmatic model for the study, for example, of the entanglement of two electrons \cite{Coe08,Man10} or of the origin of Hund's rule \cite{Sak10}. In the former case, the entanglement of low-energetic states increases with the interaction between the electrons and with the energy of the state. This is consistent with investigations in low-excited states of helium \cite{Deh12}. However, the onset of chaotic dynamics might induce qualitatively new features \cite{Tic11}. In the latter case, the spin symmetry determines the symmetry of the spatial wave function under particle exchange, leading to the effect known as the Fermi hole, describing the minimum of the triplet wave function around the origin. Still, it is not \textit{a priori} clear, which eigenenergy is energetically favorable for identical spatial configuration but different spin symmetry.
 
Extending the harmonic model, the next to leading order quartic term in the potential has so far received very little attention in the literature, e.g. Ref. \onlinecite{Pfa91,Sak06}. Yet, the enhanced complexity of the dynamics induced by this anharmonicity -- as a consequence of the loss of separability between the COM and the relative motion -- might help for a better understanding of, e.g., the entanglement of two electrons in atomic systems \cite{Maj12}. Furthermore the validity of Hund's rules in an anharmonic case has not been addressed, so far. A similar effect has been seen earlier for the quantum dot model, where the spin symmetry of the ground state depends on the applied magnetic field \cite{Din94}. 

Measurements of the electric current through a single quantum dot depending on the applied gate voltage show a specific shell structure of the ground state energies for few electrons confined in the dot \cite{Tar96,Kou01}. This shell structure supports a planar approach with a harmonic confining potential, which has extensively been studied in the past, most frequently for the two electron case  \cite{Cif04}. A detailed review on the properties and modeling of semiconductor quantum dots can be found in Ref. \onlinecite{Rei02}. The question of the relevance of the planar model has been discussed in detail in Ref. \onlinecite{Naz02}, including the effects of a static magnetic field perpendicular to the plane. 
The planar confinement generates an energy shift, which can be related to the ground state energy of a strongly confined harmonic potential in the $z$-direction, $E_z=\hbar\omega_z/2$. In addition, the Coulomb interaction is overestimated by the planar restriction. These effects can be properly incorporated in a planar approach by an appropriate rescaling justifying this model under strong confinement to the plane, $\omega_z\gg\omega_{xy}$.

The aim of this work is to shed some light on the understanding of the complex dynamics induced by a quartic potential in a two-electron quantum dot. For this purpose we have developed a quantum approach to an anharmonic two-electron quantum dot confined to a plane in analogy to a planar model for two-electron atoms \cite{Hil02,Mad08}. In Sec.  \ref{sec_theo_app} we describe our extended model of a two-electron quantum dot and clarify the underlying symmetry structure, already inherent to the non-interacting harmonic model. We show the convergence of our results and compare to available data in the literature. After a brief description of the mixed regular-chaotic structure of the classical phase space of the system in Sec. \ref{sec_classical}, we proceed to explore the role of the anharmonicity in the complexity of the quantum mechanical model in Sec. \ref{sec_eff_anharm}. Finally we conclude in Sec. \ref{sec_con}.

\section{\label{sec_theo_app}Theoretical approach}
\subsection{Physical model}
The problem we wish to describe with a minimum of approximations is a system of two interacting fermionic negative charges confined in an anharmonic potential. Each charge has an effective mass $m_*$ and the Coulomb repulsion between them is affected by the dielectric properties of the confining system manifested  through the dielectric constant $\epsilon$. Introducing modified atomic units (see Appendix \ref{sec_units}) the quantities $m_*$ and $4\pi\epsilon_0\epsilon$ can be set to unity. In the course of this work we are going to use these modified units ([a.u.]). The general form of the Hamiltonian describing our model reads as
\begin{equation}\label{eq_Ham}
H=\sum_{j=1}^2\left[-\frac{1}{2} \boldsymbol{\nabla}_j^{\,2}+V_{\rm conf}(\boldsymbol{r}_j)\right]+V_{\rm int}(r_{12}).
\end{equation}
Here $\boldsymbol{\nabla}_j$ acts only on $\boldsymbol{r}_j$ and $V_{\rm int}(r_{12})={1}/{r_{12}}$ is the interaction potential between the two charges separated by a distance $r_{12}=\left|\boldsymbol{r}_1-\boldsymbol{r}_2\right|$. For practical purposes we introduce a control parameter $\gamma$ such that $V_{\rm int}(r_{12})={\gamma}/{r_{12}}$. The anharmonicity of the quantum dot is modeled by a quartic contribution $\kappa r^4$ to the harmonic confinement $\frac12\omega_x^2\,x^2+\frac12\omega_y^2\,y^2+\frac12\omega_z^2\,z^2$. If we assume $\omega_z\gg\omega_x,\omega_y$, the dynamics in the quantum dot is strongly confined to the $xy$-plane. This is the case we consider in this contribution. From now on, we restrict the dynamics to two dimensions of configuration space, with the Cartesian positions $(x_1 , y_1)$ and $(x_2 , y_2)$ of the charges. We assume furthermore that the confinement potential is isotropic, $V_{\rm conf}(\boldsymbol{r})=\frac12 \omega^2 r^2+\kappa r^4$ ($\omega=\omega_x=\omega_y$). 

The eigenvalue problem for $\gamma\ne 0$ effectively depends on two parameters, namely $\kappa$ and $\omega$. Following experimental data we set $\omega\equiv1$ (see appendix \ref{sec_units}).  For small values of the quartic potential strength $\kappa$ the experimentally confirmed symmetry structure of the planar harmonic model is effectively not broken \cite{Tar96}. Therefore we vary the anharmonicity of the potential with the help of the parameter $\kappa$, which ranges from $0.0$ to $0.1$ for the interacting ($\gamma=1$) and non-interacting cases ($\gamma=0$).

\subsection{\label{sec_pho}Underlying symmetry structure}
To gain insight into the level structure of the full system (\ref{eq_Ham}), it is instructive to consider the energy levels and their degeneracies in the fermionic, non-interacting planar harmonic oscillator model. There are several possibilities to distinguish between symmetry classes, but our choice is made in order to resolve all degeneracies as soon as any kind of perturbation is introduced ($\kappa\ne 0$ or $\gamma \ne0$).

We start from the radial representation in COM and relative coordinates, which gives rise to the eigenbasis $|n_c,m_c,n_r,m_r\rangle$ with the principal quantum numbers $n_c$ and $n_r$ and the angular momentum quantum numbers perpendicular to the plane $m_c$ and $m_r$, for the COM and relative motion, respectively. The energies are given by $E_{c/r}=\omega(2n_{c/r}+|m_{c/r}|+1)$. These quantum numbers lose their meaning when the anharmonic potential is introduced due to the coupling of the COM and relative motion. On the other hand, the squared total angular momentum operator perpendicular to the plane $L_z^2$, the particle exchange operator $\Pi_{12}$ and the operator $\Pi_{xy}$, which interchanges the spatial coordinates $x_i$ and $y_i$, commute with the Hamiltonian (\ref{eq_Ham}). The latter is a two-dimensional parity operator, as it changes the orientation of the coordinate system. The action of the exchange and interchange operators on some function $\psi(x_1,y_1,x_2,y_2)$ in coordinate space is given 
by
\begin{align}
 \Pi_{12}\,\psi(x_1,y_1,x_2,y_2)&=\pm\psi(x_2,y_2,x_1,y_1),\\
 \Pi_{xy}\,\psi(x_1,y_1,x_2,y_2)&=\pm\psi(y_1,x_1,y_2,x_2),
\end{align}
respectively. From this and from the coordinate representation of $L_z^2$,
\begin{equation}
 L_z^2=\left(x_1 p_{y_1} - y_1 p_{x_1}+x_2 p_{y_2} - y_2 p_{x_2}\right)^2,
\end{equation}
it becomes apparent, that these three and the full Hamiltonian are mutually commuting operators. A common set of eigenstates of $L_z^2$, $\Pi_{12}$ and $\Pi_{xy}$ is defined by 
\begin{align*}
 |n_c,m_c,n_r,m_r\rangle^{\epsilon_p}&=\\
 ( |n_c,m_c,n_r,m_r\rangle&+\epsilon_p |n_c\,\text{-}m_c\,n_r\,\text{-}m_r\rangle)/\sqrt{2},
\end{align*}
with $\epsilon_p\in\lbrace+1\,\text{(even)},-1\,\text{(odd)}\rbrace$. In order to guarantee uniqueness of the representation of this eigenbasis it is necessary to impose $m_c>0$ and if $m_c=0$ then $m_r>0$. For the case with $m_c=m_r=0$, we set:
\begin{equation*}
  |n_c,m_c,n_r,m_r\rangle^{\epsilon_p\equiv1}=|n_c,m_c,n_r,m_r\rangle.
\end{equation*}
The action of the symmetry operators on this basis is given by
\begin{align*}
 L_z^2|n_c,m_c,n_r,m_r\rangle^{\epsilon_p}&=|m_c+m_r|^2&|n_c,m_c,n_r,m_r\rangle^{\epsilon_p},\\
 \Pi_{12}|n_c,m_c,n_r,m_r\rangle^{\epsilon_p}&={\rm e}^{{\rm i}\pi m_r}&|n_c,m_c,n_r,m_r\rangle^{\epsilon_p},\\
 \Pi_{xy}|n_c,m_c,n_r,m_r\rangle^{\epsilon_p}&=\epsilon_p&|n_c,m_c,n_r,m_r\rangle^{\epsilon_p},
\end{align*}
where the second identity stems from the fact, that the particle interchange does only affect the relative coordinates since it introduces a rotation by $\pi$ about the relative $z$-axes. We identify the quantum numbers:
\begin{align*}
|m_c+m_r|=m&\in\mathbb{N}_0,\\
2 (m_r\,{\rm mod}\, 2)=s&\in\lbrace0\,\text{(singlet)},\,2\,\text{(triplet)}\rbrace,
\end{align*}
where the choice of $s$ for the particle exchange operator is motivated by the numerical basis representation (see Sec. \ref{sec_bas_rep}). The principal quantum number $n$ describing the unperturbed energy levels $E_n=\omega(n+2)$ is $n=2(n_c+n_r)+|m_c|+|m_r|$. The structure of the unperturbed spectrum for the lowest lying states of total angular momentum $0$ and $1$ is shown in Table \ref{tab_deg_pxy}.

Turning on the interparticle interaction does not introduce any change for the COM motion. For the relative motion 
there is not any more a closed expression for the energy. Nevertheless two quantum numbers $\tilde{n}_r$ and $\tilde{m}_r$ can be identified using, for instance,  a WKB approach \cite{Kla98}. Therefore, the classification is still exact as long as no anharmonic interaction is present. 

For the full potential case the principal quantum number loses its meaning when avoided crossings appear in the spectrum and symmetries of energy eigenstates interchange (see Sec. \ref{sec_ac}). 

\begin{table}[t]\centering
\begin{tabular}{|r|r|r||r|r|r|r|r|r|r|r|}\hline
\multicolumn{3}{|c|}{$L_z^2$ ($m$)} & \multicolumn{4}{c|}{$0$} & \multicolumn{4}{c|}{$1$}\\\hline
\multicolumn{3}{|c|}{$\Pi_{12}$ ($s$)} & \multicolumn{2}{c|}{singlet} & \multicolumn{2}{c|}{triplet} & \multicolumn{2}{c|}{singlet} & \multicolumn{2}{c|}{triplet} \\\hline
\multicolumn{3}{|c|}{$\Pi_{xy}$ ($\epsilon_p$)} & \multicolumn{1}{c|}{$+1$} & \multicolumn{1}{c|}{$-1$} & \multicolumn{1}{c|}{$+1$} & \multicolumn{1}{c|}{$-1$} & \multicolumn{1}{c|}{$+1$} & \multicolumn{1}{c|}{$-1$} & \multicolumn{1}{c|}{$+1$} & \multicolumn{1}{c|}{$-1$} \\\hline \cline{1 - 3} \cline{1 - 3}
$n$ &   $E_n$ & $\#E_n$ & $K_n$ &  $K_n$ &  $K_n$ & $K_n$ &  $K_n$ & $K_n$ & $K_n$ &  $K_n$ \\\hline
 0 &  2 &     1 &   1 &   0 &   0 &   0 &   0 &   0 &   0 &   0 \\
 1 &  3 &     4 &   0 &   0 &   0 &   0 &   1 &   1 &   1 &   1 \\
 2 &  4 &    10 &   2 &   0 &   1 &   1 &   0 &   0 &   0 &   0 \\
 3 &  5 &    20 &   0 &   0 &   0 &   0 &   3 &   3 &   3 &   3 \\
 4 &  6 &    35 &   4 &   1 &   2 &   2 &   0 &   0 &   0 &   0 \\
 5 &  7 &    56 &   0 &   0 &   0 &   0 &   6 &   6 &   6 &   6 \\
 6 &  8 &    84 &   6 &   2 &   4 &   4 &   0 &   0 &   0 &   0 \\
 7 &  9 &   120 &   0 &   0 &   0 &   0 &  10 &  10 &  10 &  10 \\
 8 & 10 &   165 &   9 &   4 &   6 &   6 &   0 &   0 &   0 &   0 \\
 9 & 11 &   220 &   0 &   0 &   0 &   0 &  15 &  15 &  15 &  15 \\
10 & 12 &   286 &  12 &   6 &   9 &   9 &   0 &   0 &   0 &   0 \\
\vdots & \vdots & \vdots & \vdots & \vdots & \vdots & \vdots & \vdots & \vdots & \vdots & \vdots \\
39 & 41 & 11480 &   0 &   0 &   0 &   0 & 210 & 210 & 210 & 210 \\
40 & 42 & 12341 & 121 & 100 & 110 & 110 &   0 &   0 &   0 &   0 \\\hline\hline
\multicolumn{2}{|c|}{$\sum$} & 135751 & 946 &715 &  825 &  825 & 1540 & 1540 & 1540 & 1540 \\\hline
\end{tabular}
\caption[Degeneracies for $\Pi_{xy}$]{The degeneracies of the planar two-dimensional harmonic oscillator with two Fermions with respect to the symmetry operators $L_z^2$, $\Pi_{12}$ and $\Pi_{xy}$. The total degeneracy $\#E_n=\frac16 (n+1)(n+2)(n+3)$ takes into account all symmetry classes, whereas $K_n$ only counts the degeneracy within the specified symmetry class.}
\label{tab_deg_pxy}
\end{table}

\subsection{Operator representation}
All relevant physical information is contained in the spectrum of the Hamiltonian (\ref{eq_Ham}). 
One of the major complications for the diagonalization of this Hamiltonian, inherent to all numerical approaches considering few-body problems, is the treatment of Coulomb singularities. These can be rigorously regularized using a representation in the parabolic coordinates ($\mu_+,\nu_+,\mu_-,\nu_-$) defined by
\begin{equation}\label{eq_ct}
 \begin{aligned}
   \mu_+     & =\sqrt{R_++x_+},          & x_{\pm}   &=x_1\pm x_2,\\
   \nu_+     & =\sgn(y_+)\sqrt{R_+-x_+}, & y_{\pm}   &=y_1\pm y_2,\\
   \mu_-     & =\sqrt{R_-+x_-},          & R_{\pm}   &=+\sqrt{x_{\pm}^2+y_{\pm}^2},\\
   \nu_-     & =\sgn(y_-)\sqrt{R_--x_-}, & \sqrt{g}  &=R_+\;R_-\,,
\end{aligned}
\end{equation}
where $\sqrt g$ is the Jacobian of the transformation. 
Notice that $r_{12}=R_-=\mu_-^2+\nu_-^2$ is a polynomial expression of the new coordinates. Furthermore, the kinetic energy $K=-\frac12\boldsymbol{\nabla}_1^{\,2}-\frac12\boldsymbol{\nabla}_2^{\,2}$ after multiplication by the Jacobian, 
\begin{align}
\sqrt gK=&-\frac14\left[\left(\mu_-^2+\nu_-^2\right)\left(\partial_{\mu_+}^2+\partial_{\nu_+}^2\right)\right.\nonumber\\
&+\left.\left(\mu_+^2+\nu_+^2\right)\left(\partial_{\mu_-}^2+\partial_{\nu_-}^2\right)\right]\,,
\end{align}
is a polynomial expression of the parabolic coordinates and their derivatives. The same holds for all terms of the generalized eigenvalue problem (GEVP)
\begin{equation}\label{gevp}
   A|\Psi\rangle=E B|\Psi\rangle,
\end{equation}
with $A=\sqrt{g}H$ and $B=\sqrt{g}$, obtained after multiplication of the stationary Schr\"odinger equation by the Jacobian $\sqrt g$. This offers the opportunity of a representation in circular harmonic oscillator creation and annihilation operators defined by
\begin{align*}
   a_1=&\frac{1}{2}\left(\mu_+ +\partial_{\mu_+}-{\rm i}\,\nu_+ -{\rm i}\,\partial_{\nu_+}\right),\\
   a_1^\dagger=&\frac{1}{2}\left(\mu_+ -\partial_{\mu_+}+{\rm i}\,\nu_+ -{\rm i}\, \partial_{\nu_+}\right),\\
   a_2=&\frac{1}{2}\left(\mu_+ +\partial_{\mu_+}+{\rm i}\, \nu_+ +{\rm i}\, \partial_{\nu_+}\right),\\
   a_2^\dagger=&\frac{1}{2}\left(\mu_+ -\partial_{\mu_+}-{\rm i}\, \nu_+ +{\rm i}\, \partial_{\nu_+}\right),\\
   a_3=&\frac{1}{2}\left(\mu_- +\partial_{\mu_-}-{\rm i}\, \nu_- -{\rm i}\, \partial_{\nu_-}\right),\\
   a_3^\dagger=&\frac{1}{2}\left(\mu_- -\partial_{\mu_-}+{\rm i}\, \nu_- -{\rm i}\, \partial_{\nu_-}\right),\\
   a_4=&\frac{1}{2}\left(\mu_- +\partial_{\mu_-}+{\rm i}\, \nu_- +{\rm i}\, \partial_{\nu_-}\right),\\
   a_4^\dagger=&\frac{1}{2}\left(\mu_- -\partial_{\mu_-}-{\rm i}\, \nu_- +{\rm i}\, \partial_{\nu_-}\right).
\end{align*}
After normal ordering we arrive at a representation of $A$ and $B$ consisting of 2088 and 25 ordered monomials, respectively, and the maximum degree is 12. For example, we give a few terms of the regularized quartic potential:
\begin{equation}\label{eq_quart}
 H_{\rm quart}=\frac{3}{4}+\frac{7}{16} (a_1^\dagger)^3 a_1^3+\frac{5}{128}a_1 (a_2^\dagger)^3 a_3 a_4^5+\dots\,.
\end{equation}
A much simpler expression is obtained for the angular momentum $L_z=x_1 p_{y_1} - y_1 p_{x_1}+x_2 p_{y_2} - y_2 p_{x_2}$\,,
\begin{equation*}
L_z=\frac12(a_1^\dagger a_1-a_2^\dagger a_2+a_3^\dagger a_3-a_4^\dagger a_4)=\frac12(\hat{n}_1-\hat{n}_2+\hat{n}_3-\hat{n}_4)\,,
\end{equation*}
where $\hat{n}_k=a_k^\dagger a_k$ are the corresponding number operators.

\subsection{\label{sec_bas_rep}Basis representation}
Since the circular operators satisfy the usual commutation relations,
\begin{equation}
\label{commut}
[a_i,a_j]=0,\hspace{1cm}[a_i^\dagger,a_j^\dagger]=0,\hspace{1cm}[a_i,a_j^\dagger]=
\delta_{ij},
\end{equation}
for $i,j=1,2,3,4$, we can associate a harmonic oscillator with
 each pair of circular operators
$a_i^\dagger$ and $a_i$, which induces a natural
 basis set composed of tensor products of harmonic
oscillator Fock states:
\begin{equation}
\label{basis0}
|n_1n_2n_3n_4\rangle=|n_1\rangle\otimes|n_2\rangle\otimes|n_3\rangle\otimes|n_4\rangle.
\end{equation}
Each monomial element of a polynomial operator $O$ in ladder operator representation couples each basis element with exactly one element of the basis.
Two elements $|n_1n_2n_3n_4\rangle$ and $|n_1'n_2'n_3'n_4'\rangle$ of the
basis set (\ref{basis0})  are coupled or satisfy
the selection rule $\{\Delta n_1,\Delta n_2,\Delta n_3,\Delta n_4\}$, with $\Delta
n_i=n_i-n_i'$, if 
$\langle n_1n_2n_3n_4|O|n_1'n_2'n_3'n_4'\rangle\ne0$.
 For example, the selection rule defined by the monomial $a_1(a_2^\dagger)^3 a_3 a_4^5$, appearing in the representation of the quartic potential \eqref{eq_quart}, is $\lbrace  \Delta n_1, \Delta n_2, \Delta n_3, \Delta n_4\rbrace=\lbrace-1,3,-1,-5\rbrace$. The operator $A$ in (\ref{gevp}) defines 171 selection rules, while the Jacobian operator $B$ has 9 selection rules. The only selection rule of the angular momentum $L_z$ is $\lbrace 0,0,0,0\rbrace$ which trivially implies that the basis elements $|n_1n_2n_3n_4\rangle$ are eigenvectors of $L_z$ (with eigenvalue $\frac12(n_1-n_2+n_3-n_4)$).
For a given selection rule 
$\Delta\nn= \{ \Delta n_1,\Delta n_2,\Delta n_3,\Delta n_4\}$,  the matrix elements $\langle \nn+\Delta\nn|A|\nn\rangle$
and $\langle \nn+\Delta\nn|B|\nn\rangle$, with 
 $|\nn\rangle=|n_1n_2n_3n_4\rangle$ and
$|\nn+\Delta \nn\rangle=
|n_1+\Delta n_1~n_2+\Delta n_2~n_3+\Delta n_3~n_4+\Delta n_4~\rangle$, involve
square roots of integer numbers and 
 depend only on $n_1,~n_2,~n_3$ and $n_4$. For example, the matrix
element of the operator $A$ for the selection rule
$\Delta \boldsymbol{n}=\lbrace-1,3,-1,-5\rbrace$ reads
\begin{eqnarray*}
\langle \nn +\Delta\nn|A|\nn\rangle&=&\sqrt{n_1(n_2+1)(n_2+2)(n_2+3)}\\
&&\times\sqrt{n_3(n_4-3)(n_4-2)(n_4-1)n_4}
\\
&&\times\left(\frac{5}{128}+\frac{5}{256}(n_1-1)+\frac{5}{512}n_2\right)\,.
\end{eqnarray*}
The parabolic transformation introduces a four-times-layer of the coordinate space resulting in unphysical symmetries. This must be compensated by a restriction of the allowed basis vectors. Indeed, only even values of  $n_1-n_2$ and $n_3-n_4$ have a physical meaning.

Since a particle exchange can be identified 
with a rotation in the parabolic coordinate subspace $(\mu_-,\nu_-)$ the basis (\ref{basis0}) is an eigenbasis of $\Pi_{12}$. This follows from the identity 
\begin{align*}
   \Pi_{12}\left(\mu_+,\nu_+,\mu_-,\nu_-\right)=&\left(\mu_+,\nu_+,\pm\nu_-,\mp\mu_-\right)\\
   \equiv& e^{\pm {\rm i}\frac{\pi}{2} L_-}\left(\mu_+,\nu_+,\mu_-,\nu_-\right),
\end{align*}
with $L_-=-{\rm i}\left(\mu_-\partial_{\nu_-}-\nu_-\partial_{\mu_-}\right)=a_3^\dagger a_3-a_4^\dagger a_4$,
which equivalently can be written as
\begin{align*}
   \Pi_{12}\mid n_1n_2n_3n_4\rangle=&e^{\pm {\rm i}\frac{\pi}{2}L_-}\mid n_1n_2n_3n_4\rangle\\
   =&e^{\pm {\rm i}\frac{\left(n_3-n_4\right)\pi}{2}}\mid n_1n_2n_3n_4\rangle\,.
\end{align*}
The two symmetry classes defined by $\Pi_{12}$ are thus identified with the quantum number $s=0$ (singlet states) or $s=2$ (triplet states) such that $n_3-n_4\equiv s\pmod 4$.

The action of the coordinate exchange operator $\Pi_{xy}$ on an element of the basis (\ref{basis0}) is given by
\begin{equation}
 \Pi_{xy}|n_1n_2n_3n_4\rangle=|n_2n_1n_4n_3\rangle\,.
\end{equation}
This can be easily seen from the coordinate representation of the operators $\hat{n}_k$, $k=1,2,3,4$:\begin{equation*}
 \hat n_k=\frac12\left(R_\pm-R_\pm\nabla_\pm^2-1\right)+{\rm i}(-1)^k\left(x_\pm\partial_{y_\pm}-y_\pm\partial_{x_\pm}\right)\,,
\end{equation*}
with $\nabla^2_\pm=\partial_{x_\pm}^2+\partial_{y_\pm}^2$. Plus signs correspond to $k\in\lbrace1,2\rbrace$, while minus signs correspond to $k\in\lbrace3,4\rbrace$.

Summarizing, a common eigenbasis of the operators $L_z^2$, $\Pi_{12}$ and $\Pi_{xy}$ is defined by
\begin{equation}
 \label{eq:simmbasis}
|n_1n_2n_3n_4\rangle^{\epsilon_p}=\frac{1}{\sqrt2}\left(|n_1n_2n_3n_4\rangle+\epsilon_p|n_2n_1n_4n_3\rangle\right),
\end{equation}
and the associated quantum numbers are
\begin{align*}
 m&=\frac12 \left|n_1-n_2+n_3-n_4\right|,\\
 s&=(n_3-n_4)\pmod 4,\\
 \epsilon_p&=\pm1\,,
\end{align*}
respectively.

To ensure a unique representation of the basis vectors we impose restrictions on the basis set resulting in three cases:
\begin{align*}
 1)&&n_3 & > n_4;~ n_1, n_2 \text{ arbitrary},\\
 2)&&n_3 & = n_4 \text{ and } n_1 > n_2,\\
 3)&&n_3 & = n_4 \text{ and } n_1 = n_2,\\
&&&\text{but then } |n_1n_2n_3n_4\rangle^{\epsilon_p\equiv1}=|n_1n_2n_3n_4\rangle.
\end{align*}

\begin{table*}[t]\centering
 \begin{tabular}{|r|r|r|r|r|}\hline
 $E$           &   $\langle{\rm d}H/{\rm d}\alpha\rangle$      &   $\langle R_{\rm COM} \rangle$ (numerical) & $\langle R_{\rm COM} \rangle$ (analytical) & $\langle r_{\rm rel}\rangle$ (numerical) \\\hline
 2.999 999 999 999 999 &   0.000 000 000 000 049 &   0.626 657 068 657 750 & 0.626 657 068 657 750 &  1.636 801 341 900 274\\
 5.000 000 000 000 005 &  -0.000 000 000 000 008 &   1.096 649 870 151 057 & 1.096 649 870 151 063 &  1.636 801 341 900 264\\
 7.000 000 000 000 039 &  -0.000 000 000 000 012 &   1.419 769 921 177 681 & 1.419 769 921 177 715 &  1.636 801 341 900 233\\
 9.000 000 000 000 684 &  -0.000 000 000 000 206 &   1.681 692 992 842 792 & 1.681 692 992 843 259 &  1.636 801 341 899 816\\
11.000 000 000 001 018 &   0.000 000 000 000 223 &   1.907 854 079 694 065 & 1.907 854 079 694 655 &  1.636 801 341 899 770\\
13.000 000 000 000 819 &   0.000 000 000 000 113 &   2.109 832 797 669 925 & 2.109 832 797 670 389 &  1.636 801 341 899 910\\
15.000 000 000 000 076 &  -0.000 000 000 000 100 &   2.294 050 646 216 108 & 2.294 050 646 216 114 &  1.636 801 341 900 266\\
17.000 000 000 000 028 &   0.000 000 000 000 184 &   2.464 507 391 773 477 & 2.464 507 391 773 425 &  1.636 801 341 900 298\\
19.000 000 000 000 298 &   0.000 000 000 050 815 &   2.623 894 283 580 835 & 2.623 894 283 588 085 &  1.636 801 341 895 081\\
21.000 000 000 132 665 &  -0.000 000 022 123 264 &   2.774 124 916 536 298 & 2.774 124 922 956 349 &  1.636 801 338 158 184\\
23.000 000 005 786 998 &   0.000 000 768 108 903 &   2.916 617 106 250 236 & 2.916 617 677 766 815 &  1.636 801 016 875 796\\
25.000 000 010 704 806 &  -0.000 002 561 041 945 &   3.052 460 181 107 163 & 3.052 458 493 794 680 &  1.636 802 260 485 613\\
27.000 000 121 766 810 &   0.000 304 331 105 631 &   3.182 573 077 654 419 & 3.182 500 734 139 942 &  1.636 838 163 687 941\\
29.000 001 160 891 170 &  -0.000 115 325 875 665 &   3.307 270 472 895 011 & 3.307 429 480 650 727 &  1.636 722 556 978 399\\
31.000 002 459 204 911 &   0.000 453 485 791 883 &   3.427 756 436 953 774 & 3.427 804 635 093 639 &  1.636 775 103 185 396\\
32.999 974 115 691 302 &  -0.000 709 199 538 930 &   3.548 166 414 577 438 & 3.544 090 785 080 888 &  1.638 653 874 804 282\\
35.000 074 767 878 694 &  -0.001 304 748 848 729 &   3.658 326 733 913 608 & 3.656 678 487 565 035 &  1.637 404 037 276 759\\
37.000 475 686 981 808 &  -0.003 970 214 278 542 &   3.782 104 880 243 939 & 3.765 899 804 871 662 &  1.643 074 413 780 297\\
39.002 630 249 149 078 &  -0.058 636 365 021 995 &   3.918 894 728 959 541 & 3.872 039 883 745 950 &  1.653 311 492 871 477\\\hline
 \end{tabular}
\caption[Comparison to analytical results]{Numerically calculated eigenenergies $E$ and expectation values  $\langle{\rm d}H/{\rm d}\alpha\rangle$, $\langle R_{\rm COM} \rangle$, and $\langle r_{\rm rel}\rangle$  of singlet states with  even parity and vanishing angular momentum for $\omega=1$, $\kappa=0$, and $\gamma=1$.  For this special choice of the parameters $\omega$, $\kappa$, and $\gamma$ all odd integer values represent exact eigenenergies and the radial expectation value equals $\langle r_{\rm rel}\rangle=2(2+\sqrt{2\pi})/(3+\sqrt{2\pi})=1.636\,801\,341\,900\,272$ (rounded to the last digit). 
}
\label{tab_harmonic}
\end{table*}
\subsection{Observables and expectation values}
The expectation value $\langle O\rangle$ of an observable $O$ for a general state $|\Psi\rangle$ is evaluated with the help of the expression $\langle \Psi|BO|\Psi\rangle$. The Jacobian matrix $B$ must be included due to the orthogonality relation for the eigenstates $|\Psi_i\rangle$ of the GEVP (\ref{gevp}),
$\langle\Psi_i|B|\Psi_j\rangle=\delta_{ij}$.
Knowing the expansion coefficients of $|\Psi\rangle$ in the symmetrized basis $|\nn\rangle^{\epsilon_p}$  (Eq. (\ref{eq:simmbasis})) the expectation value $\langle O\rangle$ can be readily obtained from the matrix representation of 
$BO$ in this basis. This is a simple task in the case that  $BO$ is a polynomial function of  the parabolic coordinates and, therefore, has a finite representation in creation and annihilation operators. Operators satisfying such property are the relative distance $r_{\rm rel}$ and the COM position $R_{\rm COM}$.
For example,
\begin{equation*}
 B R_{\rm COM}=\frac12 B R_+=\frac{1}{16}\left(\mu_+^2+\nu_+^2\right)^2\left(\mu_-^2+\nu_-^2\right)
\end{equation*}
is clearly a polynomial expression of the parabolic coordinates. Its ordered representation in ladder operators contains 70 terms and defines 15 selection rules. The cosine of the angle $\varphi_{12}$ between the electron radii can be estimated from a combination of radial expectation values:
\begin{equation*}
 \cos\varphi_{12}\approx\frac{\langle \boldsymbol r_1 \cdot\boldsymbol r_2\rangle}{\langle|\boldsymbol r_1|\rangle\langle|\boldsymbol r_2|\rangle}\approx\frac{\langle R_+^2\rangle -\langle R_-^2\rangle}{4\sqrt{\langle\boldsymbol r_1\,^2\rangle}\sqrt{\langle\boldsymbol r_2\,^2\rangle}}.
\end{equation*}
The structure of our code allows for separate calculation of the different parts of the Hamiltonian, the kinetic term $T$, and the three potential terms involved. It is easy to show, that our system satisfies a generalized virial theorem,
\begin{equation*}
 2\langle T \rangle = 2 \langle V_{\rm harmonic} \rangle + 4 \langle V_{\rm quartic} \rangle - \langle V_{\rm Coulomb} \rangle\,,
\end{equation*}
 where $\langle \cdot\rangle$ denotes the expectation value in an eigenstate.
Equivalently this may be changed to an expression for the energy depending only on the potential terms of the system:
\begin{equation}\label{eq_virial}
 E=2 \langle V_{\rm harmonic} \rangle + 3 \langle V_{\rm quartic} \rangle + \frac12\langle V_{\rm Coulomb} \rangle.
\end{equation}
Our numerical results agree with the virial theorem \eqref{eq_virial} up to the full accuracy of the eigenenergies.

\subsection{\label{sec_num_app} Convergence of the method}

The symmetrized basis (\ref{eq:simmbasis}) allows an analytic, exact representation of the GEVP \eqref{gevp}. For practical purposes this basis has to be truncated. We use the truncation criterion
\begin{equation}
 n_1+n_2+n_3+n_4\le n_{\rm base},
\end{equation}
for a given integer number $n_{\rm base}$. 
Typical values of $n_{\rm base}$ in our calculations go up to 140, which translate into basis dimensions up to 16530. 
A Krylov subspace method for symmetric matrices -- which is generally known as the Lanczos algorithm \cite{lanczos1} -- exploits the band structure of these matrices and efficiently calculates the largest eigenvalues of the problem. We take advantage of this property by shifting the energy and solving the inverse problem, which leads to a well converged spectrum around the shifted energy \cite{Eri80,krug1}. 

In order to verify the convergence of the eigenvalues obtained we introduce the scaling transformation mediated by the unitary operator $P_\alpha=\exp[-\frac{\rm i}{2}(\boldsymbol r\cdot\boldsymbol p+\boldsymbol p\cdot\boldsymbol r)\log\alpha]$, with $\alpha$ a real scaling factor. Position and momentum are transformed according to $\boldsymbol{r}\rightarrow \alpha\boldsymbol{r}$ and $\boldsymbol{p}\rightarrow \boldsymbol{p}/\alpha$.
The physical properties are not altered by this unitary transformation. In particular the exact eigenvalues are invariant under this transformation,
\begin{equation}
\label{eq:dealpha}
 \frac{{\rm d} E_\alpha}{{\rm d}\alpha}=\langle E_\alpha|\frac{{\rm d}H}{{\rm d}\alpha}|E_\alpha\rangle=0.
\end{equation}
The truncation of the basis leads however to an $\alpha$-dependency of eigenenergies. The expectation value (\ref{eq:dealpha}) does not vanish anymore, but it is small for well-converged eigenvalues.  $\alpha$ can thus be treated as a variational parameter. For convergence of an eigenvalue $E_\alpha$ we demand that $\langle E_\alpha|{\rm d}H/{\rm d}\alpha|E_\alpha\rangle\le 0.1$. With this condition we typically obtain at least three figures of the eigenvalues converged.\\ 
This is illustrated in the following with the interacting harmonic case $\kappa=0$ and with the full potential case for $\omega=1$, $\kappa=0.1$ and $\gamma=1$.

\begin{table*}[t]\centering
 \begin{tabular}{|r|r|r|r|r|r|r|}
  \multicolumn{7}{c}{Lowest states:}\\\hline
  spin & parity & $\langle E\rangle$           &   $\langle{\rm d}H/{\rm d}\alpha\rangle$      &   $\langle R_{\rm COM} \rangle$  & $\langle r_{\rm rel}\rangle$ & $\langle\cos{\varphi_{12}}\rangle$\\\hline
  singlet & even & 3.445 300 210 741 99 & $< 10^{-13}$ & 0.547 213 052 020 00 & 1.410 760 235 037 41 & -0.217 712 873 312 13 \\
  singlet & odd  & 8.374 386 408 855 55 & $< 10^{-12}$ & 0.953 968 643 580 06 & 1.976 361 508 431 29 & -0.033 320 618 273 03 \\
  triplet & even & 5.745 411 335 545 50 & $< 10^{-13}$ & 0.778 149 902 559 62 & 1.673 248 129 799 30 & -0.066 974 875 946 45 \\
  triplet & odd  & 5.526 039 089 152 97 & $< 10^{-13}$ & 0.802 685 965 113 66 & 1.722 064 561 651 52 & -0.064 598 448 422 64\\\hline
  \multicolumn{7}{c}{50th excited states:}\\\hline
  spin & parity & $\langle E\rangle$           &   $\langle{\rm d}H/{\rm d}\alpha\rangle$      &   $\langle R_{\rm COM} \rangle$  & $\langle r_{\rm rel}\rangle$ & $\langle\cos{\varphi_{12}}\rangle$\\\hline
  singlet & even & 23.889 095 236 563 & $< 10^{-12}$ & 1.357 034 558 & 2.799 114 48 & -0.034 186 244 183 \\
  singlet & odd  & 31.288 073 487 & $< 10^{-11}$ & 1.494 635 & 3.035 1 & -0.015 942 \\
  triplet & even & 27.856 833 862 282 & $< 10^{-12}$ & 1.246 927 703 940 & 2.427 175 802 70 & {\color{white}-}0.026 987 470 565 9 \\
  triplet & odd  & 27.269 244 649 677 & $< 10^{-12}$ & 1.424 229 868 91 & 2.915 846 906 8 & -0.024 623 600 048 \\\hline
 \end{tabular}
\caption[Ground states of the four symmetry classes]{Zero angular momentum eigenenergy $E$ and expectation values $\langle{\rm d}H/{\rm d}\alpha\rangle$, $\langle R_{\rm COM}\rangle$, $\langle r_{\rm rel}\rangle$, and $\langle\cos{\varphi_{12}}\rangle$ of the lowest state (upper part of the table) and of the 50th excited state (lower part of the table) in each symmetry class. The expectation values were calculated from the operator representation by the same method as the Hamiltonian was constructed. We round by the last digit, which coincides for at least two different values of $\alpha$ and two different values of $n_{\rm base}$.}
\label{tab_groundstate}
 \end{table*}

\subsubsection{Harmonic case}
There exist analytical solutions for the case without quartic potential ($\omega=1$, $\gamma=1$, $\kappa=0$) \cite{Tau94a}, which in general only appear for very special combinations of the harmonic frequency $\omega$, non-integer values of $\gamma$ and an external magnetic field perpendicular to the plane applied to the quantum dot \footnote{Notice that such a magnetic field modifies the harmonic frequency and introduces a linear term in $L_z$ in the Hamiltonian.}. 

For example, for singlet spin symmetry, even parity and vanishing angular momentum the analytical expression for one of the energies of the relative motion is $E_{\rm rel}=2$  and the associated radial expectation value is  $\langle r_{\rm rel}\rangle=2(2+\sqrt{2\pi})/(3+\sqrt{2\pi})$. In combination with the solutions of the COM motion the total energy reads $E=E_{\rm rel}+E_{\rm COM}= (2\,n+1)$, with $n\in \mathbb{N}$.
We compare our results for an optimal choice of $\alpha=0.2$ and a basis size of $n=6370$. For the best converged, low-lying values we obtain results in accordance with the analytical results up to numerical accuracy of 15 digits. For the worst converged values fulfilling our criteria $|\frac{{\rm d}}{{\rm d}\alpha} E_\alpha|\le 0.1$ we still obtain four valuable digits for the eigenenergies and two for the radial expectation values (see Table \ref{tab_harmonic}).

\subsubsection{Full potential case}
We consider the full Hamiltonian (\ref{eq_Ham}) with parameters $\omega=1$, $\kappa=0.1$ and $\gamma=1$.  In this case there are no analytic solutions. The convergence of zero angular momentum eigenvalues is analyzed through their dependence on the parameters $\alpha\in\lbrace0.1,0.15,0.2,0.25\rbrace$ and $n_{\rm base}\in\lbrace80,90,100,110,120,130\rbrace $. The basis size ranges from 2560 to 13379. 
For each of these parameters the eigenvalues and the associated expectation values $\langle dH/d\alpha\rangle$, $\langle R_{\rm COM}\rangle$, $\langle r_{\rm rel}\rangle$ and $\langle\cos{\varphi_{12}}\rangle$ have been calculated. We use a coincidence criterion to establish the convergence
of these quantities: these are converged if they are obtained for at least two different values of $\alpha$ and two different values of $n_{\rm base}$. 
The number of coincident significant digits in this process provides the accuracy of the result.
The lowest eigenvalue and the 50th excited state of the four symmetry classes and their respective expectation values are summarized in Table \ref{tab_groundstate}. In this numerical experiment, eigenvalues satisfying $\langle |{\rm d}H/{\rm d}\alpha| \rangle\le 0.1$ typically exhibit at least three converged digits.

\section{Role of the anharmonicity for the complexity of the system}
\subsection{\label{sec_classical}Classical effects}

It is a remarkable property of the system \eqref{eq_Ham}, that only the case with $\kappa\ne0$ and $\gamma\ne0$ leads to a chaotic classical dynamics, which stems from the fact that the system separates in two different ways. It separates into a system of two independent particles as long as $\gamma=0$ and into COM and relative motion as long as $\kappa=0$. In both cases the angular momentum and the energy in the subsystems are preserved. The system with originally four degrees of freedom therefore has four constants of motion and the underlying dynamics is integrable. Consequently we will focus on the full potential case in our classical analysis and show some signatures of the chaotic dynamics. The total energy and the total angular momentum remain constants of motion and the phase space is effectively a six-dimensional space. Except for special cases, where the motion is further confined by the choice of the initial conditions, the dynamics cannot reasonably be visualized by Poincar\'e's surfaces of 
section. Alternatively, insight into the complexity of the system can be gained by an appropriate analysis of the dominant frequencies of the trajectories \cite{Las93}. Such frequencies can be identified with the frequencies related to the largest weight in the Fourier transform of properly selected dynamical quantities. Associated to each degree of freedom there is a dominant frequency, which coincides with the fundamental frequency for regular dynamics. Therefore, compared with the method relying on the Poincar\'e surfaces of section, the frequency analysis is more appropriate for highly dimensional systems. Its applications include investigations on the Stark-Quadratic-Zeeman problem \cite{Cor08}, the two-dimensional standard map \cite{Las92}, different multidimensional systems \cite{Las93}, and the study of the stability of the solar system \cite{Las90}.

The classical Hamilton function in COM and relative coordinates is our basis for the numerical calculation of trajectories. The equations of motion are integrated with the widely used leapfrog method and the convergence of our results is tested with a fourth order Runge-Kutta-Nystr\"om algorithm \cite{San94}. We consider trajectories in the time interval $[t_0,t_0+T]$ with $t_0=100$ and $T=400$. The energy and the angular momentum is conserved with a relative error of the order $10^{-5}$ or better. For each degree of freedom we consider the combination $x_j+{\rm i}p_j$ of the coordinate and momentum, respectively. This quantity is multiplied by the widely used Hanning Filter \cite{Las93} to avoid effects at the edges of the time interval. A discrete Fourier analysis on the above given time interval provides the frequencies with an accuracy $\Delta f=1/T$ limited by the finite time interval. The exact frequency value at the peak is found by an iterative scheme in the vicinity of the highest values of the 
discrete Fourier coefficient (golden section search). In each iteration the Fourier transform is obtained by straightforward quadrature of the Fourier integral \cite{Las93}.

For our analysis we choose initial conditions such that one electron starts at rest from a point on the $x$-axis, while the other electron starts from equally distributed positions on the circle with radius one and momentum pointing outwards. Then the total angular momentum naturally vanishes and we investigate different energy regimes.

\begin{figure}[t]
  \includegraphics[width=\columnwidth]{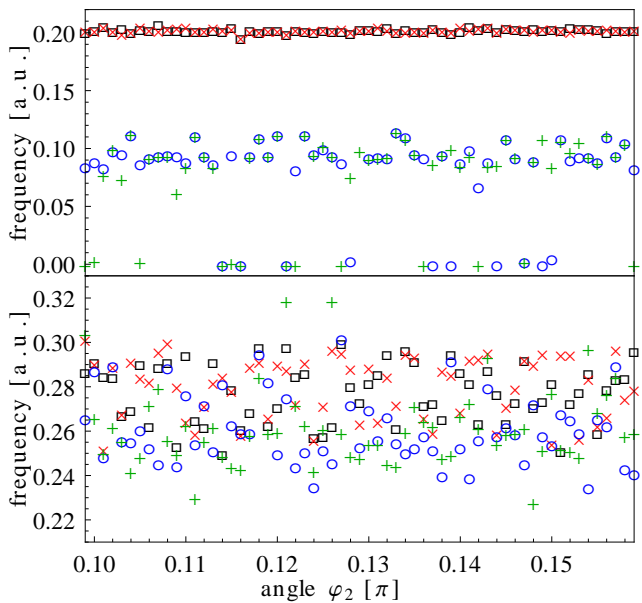} 
 \caption{\label{fig_freq} (Color online) Fundamental frequencies of the planar classical dot with $\omega=1$, $\gamma=1$ and $\kappa=0.1$ and vanishing angular momentum. Depicted are the frequencies associated to the relative motion in $x$- (circles) and $y$- direction (pluses), and to the COM motion in $x$- (squares) and $y$- direction (crosses), respectively. A clearly regular motion is recognized in the low energy regime (upper panel: $E=3$), while for the higher energy a mixed regular-chaotic dynamics (lower panel: $E=15$) is observed. The initial conditions are chosen, such that one particle starts static from $x_1=0.5$ and $y_1=0$, while the second starts from equally distributed positions on a circle of radius $r_2=1$ at the angle $\varphi_2$ with momentum pointing outwards with varying absolute value to compensate the difference in Coulomb energy.}
\end{figure}

The regularity of the system of interacting particles in the low energy regime is characterized by the smooth behavior of the dominant frequencies with respect to the initial conditions (upper panel of Fig. \ref{fig_freq}). In this case the center of mass is subject to approximately harmonic oscillations and the associated frequencies (squares and crosses) are constant and mostly degenerate. The frequencies of the relative motion (circles and pluses) change only little or can even be constant in some intervals. The dynamics is thus confined to regular nearly harmonic islands.

With increasing importance of the anharmonic potential for higher energies the dynamics is mixed regular-chaotic. This is intuitively clear from the symmetry properties of the system and can be verified in terms of the analysis of the fundamental frequencies and their sensitivity to initial conditions. A typical scenario for the frequencies in this case is shown in the lower panel of  Fig. \ref{fig_freq}. An increase in energy leads to a discontinuous behavior of the frequencies and results in a complete lifting of the degeneracy of the two frequencies of the COM motion, which is a consequence of the coupling of the COM and the relative motion.

\subsection{\label{sec_eff_anharm}Quantum effects}
Neither the combination of Coulomb interaction and harmonic potential, nor the combination of harmonic and anharmonic potential, led to chaotic classical dynamics. In the former case the relative and COM motions are separable, while in the latter case the two particles are independent. Only a coupling of these motions produces a significant impact on the complexity of the anharmonic two-electron quantum dot problem. 
In this section we investigate the effect of the anharmonicity on the complexity of the quantum system. For that purpose we rely on tools that are all related to the universal predictions of random matrix theory (RMT). These include the next-neighbor-spacing distribution $P(s)$ and the distribution $P_{\rm ac}(c)$ of the energy gaps $c$ of the avoided crossings which appear by slow variation of the parameter $\kappa$. While the universality of RMT predictions has been confirmed by several experimental \cite{Sto00}, semiclassical and numerical results for systems with underlying chaotic classical dynamics \cite{haake}, its implications for systems with mixed regular-chaotic phase space -- such as our system -- are still subject of research and not entirely understood. In the latter case, the next-neighbor-spacing distribution $P(s)$ can, with some exceptions \cite{Pro98}, be better described by the phenomenological Brody-distribution \cite{Bro81} than by the physically motivated Berry-Robnik-
distribution \cite{Ber84}. A remedy is found by extending the idea underlying the Berry-Robnik distribution, which is to split the classical phase space into distinct regular and chaotic regimes. The effects of dynamical tunneling \cite{Dav81} and chaos assisted tunneling \cite{Tom94} connect these two classically distinct regimes in the quantum regime and already improve the results \cite{Bae11,Pod07}. Recent additional achievements considering the effects of flooding \cite{Rud12} appear to complete the discussion, at least for the next-neighbor spacing distribution. Nevertheless a very thorough analysis of the classical phase space is necessary to obtain an \textit{ab initio} description of this distribution. To our knowledge, this has only been performed for one-dimensional systems. The distribution $P_{\rm ac}(c)$ is, in the mixed case, the sum of a $\delta$-peak, representing real crossings in the regular regime, and a normal distribution for the avoided crossings in the chaotic regime \cite{Yan93}
.\\
The subtlety of the symmetry and their corresponding ensembles, which is fundamental for random matrix theory, will not be further discussed here. The system under consideration belongs to the Gaussian orthogonal ensemble, as it is symmetric under time-reversal and under rotations, furthermore the Hamiltonian is real symmetric.\\
Our analysis will focus on the dependence on the anharmonicity of $P(s)$, $P_{\rm ac}(c)$ and a measure of the localization of eigenstates.
The results presented in the following were derived from data collected in different runs with the numerical method described above. We limit our calculations to zero angular momentum for all four symmetry classes (singlet/triplet spin symmetry and even/odd parity). The basis size was determined by the parameter $n_{\rm base}=130$, which leads to basis sizes of approximately 12\,000 basis vectors, where exact numbers depend on the symmetry class. Each calculation supplied more than 1000 well converged eigenvalues. The quartic potential strength was in general varied from $\kappa=0.0$ to $\kappa=0.1$ in steps of $\Delta\kappa=10^{-2}$. For the detection of the avoided crossings the step size was decreased to $\Delta\kappa=10^{-5}$ to enhance the resolution of very narrow avoided crossings.\\ 
The universal predictions of random matrix theory can only be confirmed if the characteristic spectral properties of a system are brought to some general footing. This procedure is called ``unfolding'' and different methods are in use. The common two steps for all methods are the following: (i) The level density is smoothed by a simple fit or semiclassical analysis and (ii) a new set of energies is derived from the smoothed level density, in a way, that the mean level density is normalized to unity. We fit the cumulated level density by a cubic polynomial (i) and take the value of the cumulated smooth level density evaluated at the former eigenenergy to be the new energy (ii). Alternative methods \cite{haake} produce qualitatively identical results.

\subsubsection{\label{sec_nnd}Nearest-neighbor-distribution}
The universality of chaotic properties in quantum mechanical systems has been shown for many examples involving the statistics of the separations of neighboring energy levels. Once the limits for regular (Poisson distribution) and purely chaotic (Wigner distribution) were observed in different experimental and numerical studies, the interest in situations with a mixed classical phase space arose and tools to describe the smooth transition between these regimes were developed. If there are any symmetries left in the problem, the next-neighbor-distance distribution will most likely be a Poisson distribution. Otherwise, the level statistics of an underlying mixed phase space will exhibit a level repulsion which, however, is not as pronounced as in the pure chaotic case. Alternatives for modeling these statistics are the Brody-distribution \cite{Bro81} characterized by the parameter $\beta$, 
\begin{align}
   P_{\rm Brody}(s)&=(\beta+1)a s^\beta \exp(-as^{\beta+1})\,,\nonumber
\end{align}
with $a=\Gamma\left(\frac{\beta+2}{\beta+1}\right)^{\beta+1}$;
the Berry-Robnik distribution \cite{Ber84} characterized by the parameter $\rho_{\rm BR}$, 
\begin{align}
  P_{\rm BR}(s)&=\left[\left(1-\rho_{\rm BR}\right)^2{\rm erfc}(\frac{\sqrt{\pi}}{2}\rho_{\rm BR} s)+\right.\nonumber\\
	      &+\left(2\rho_{\rm BR}(1-\rho_{\rm BR})+\frac{\pi}{2}\rho_{\rm BR}^3s\right)\times\nonumber\\
	      &\times\left.e^{-\frac{\pi}{4}\rho_{\rm BR}^2s^2}\right]e^{-(1-\rho_{\rm BR})s}\,;\nonumber
\end{align}
and the extension of the latter proposed by Podolskiy and Narimanov \cite{Pod07,Pod03} characterized by the parameters $\rho$ and $V_{RC}^0$,
\begin{align}
  P_{\rm PN}(s)&=\left[\left(1-\rho\right)^2F\left(\frac{s}{(V_{RC}^0)^2}\right){\rm erfc}(\frac{\sqrt{\pi}}{2}\rho s)\right.+\nonumber\\ 
  &\left(2\rho(1-\rho)F\left(\frac{s}{V_{RC}^0}\right)+\frac{\pi}{2}\rho^3s\right)\times\nonumber\\
  &\left.\times e^{-\frac{\pi}{4}\rho^2s^2}\right]e^{-(1-\rho)s},
\end{align}
with 
\begin{equation*}
 F(x)=1-\frac{1-\sqrt{\frac{\pi}{2}}x}{e^x-x}\,.
\end{equation*}

\begin{figure}[t]
 \includegraphics[width=\columnwidth]{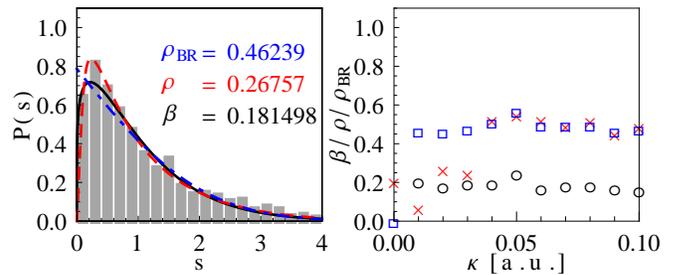}
 \caption{\label{fig_brody} (Color online) Left: The numerical level spacing distribution (gray bars) for singlet states of odd parity is shown together with the fitted Brody distribution (solid line), the fitted Berry-Robnik distribution (dot-dashed line) and the fitted Podolskiy-Narimanov distribution (dashed line) for the case with $\kappa=0.02$. Right: Brody parameter $\beta$ (circles), Berry-Robnik parameter $\rho_{BR}$ (squares) and Podolskiy-Narimanov parameter $\rho$ (crosses) singlet states of odd parity for several values of $\kappa\in\left[0,0.1\right]$.}
\end{figure}

The Brody parameter $\beta$ describes the transition from regular ($\beta=0$) to chaotic ($\beta=1$) behavior. It must be noted, that the Brody parameter $\beta$ lacks a quantitative physical meaning, but describes the transition merely qualitatively. The Berry-Robnik distribution is characterized by the parameter $\rho_{\rm BR}$, which is the ratio of the chaotic to the total phase space volume and is a purely classical property of the system. The generalization of the Berry-Robnik distribution by Podolsky and Narimanov include perturbative quantum corrections through the parameter $V_{RC}^0$ \cite{Pod07}, which describes the tunneling between regular and chaotic regions and chaos assisted tunneling between regular regions via the chaotic sea.

We have considered all states from the principal quantum numbers $n=1$ through to $n=40$, in the unperturbed system, for all symmetries, leading to a specific number of eigenenergies according to Table \ref{tab_deg_pxy}. In Fig. \ref{fig_brody} we show the results of the statistical analysis for the singlet spin symmetry odd parity case performed for 715 eigenvalues ($\alpha=0.2,~n_{\rm base}=130,~n_{\rm tot}=11\,168$). In the left panel our numerical data, for the specific case with $\kappa=0.02$, is shown as gray bars, while the different fits according to the previously described distributions are given by the solid, dashed and dot-dashed lines. For the Berry-Robnik distribution we fitted the parameter $\rho_{\rm BR}$. The fit for the numerical data using the Brody distribution is in general better than the fit by the Berry-Robnik distribution, but has the disadvantage of an unphysical fit parameter. The Podolskiy-Narimanov distribution improves the quality of the fits, naturally, as a second fit 
parameter $V_{RC}^0$ is introduced. For a quantum dot with interacting electrons ($\gamma=1$) without quartic term ($\kappa=0$) we analyzed the spectrum of the radial equation of the relative motion. Being still too close to the pathological harmonic oscillator case no proper fit could be performed.

For the full potential the Brody-parameter acquires an intermediate value of $\beta\approx0.45$ ($\beta\approx0.2$) for the singlet even parity case (other cases). Variations of the strength $\kappa$ of the quartic potential leads for $\kappa\in[0.01,0.1]$ only to small variations of $\beta$ (see right panel of Fig. \ref{fig_brody}). The parameter $\rho_{BR}$ exhibits a similar behavior however its value is shifted with respect to $\beta$ by an approximately constant value 0.3. This stems from the similarity of the two distributions with this special shift in the parameters. The parameter $\rho$ of the Podolskiy-Narimanov distribution tends to be close to one of the two previous parameters. The quantum mechanical coupling parameter $V_{RC}^0$ lies between $0.1$ and $0.45$.

Simple algebraic considerations show that the strength of the quartic potential only shifts the energy regimes for which the anharmonic effects become important. Therefore, as long as the analysis includes a regime wide enough in energy the effect of the anharmonicity can be measured for any non vanishing value of $\kappa$ and should not depend strongly on it. Performing the statistical analysis for a lower number of eigenenergies, cutting off at a specific principal quantum number $n<40$ leads to a decrease of the chaoticity parameters for $\kappa<0.05$, while the results for $\kappa\ge0.05$ remain unaltered.

\subsubsection{\label{sec_ac}Avoided crossings}

\begin{figure}[t]
 \includegraphics[width=\columnwidth]{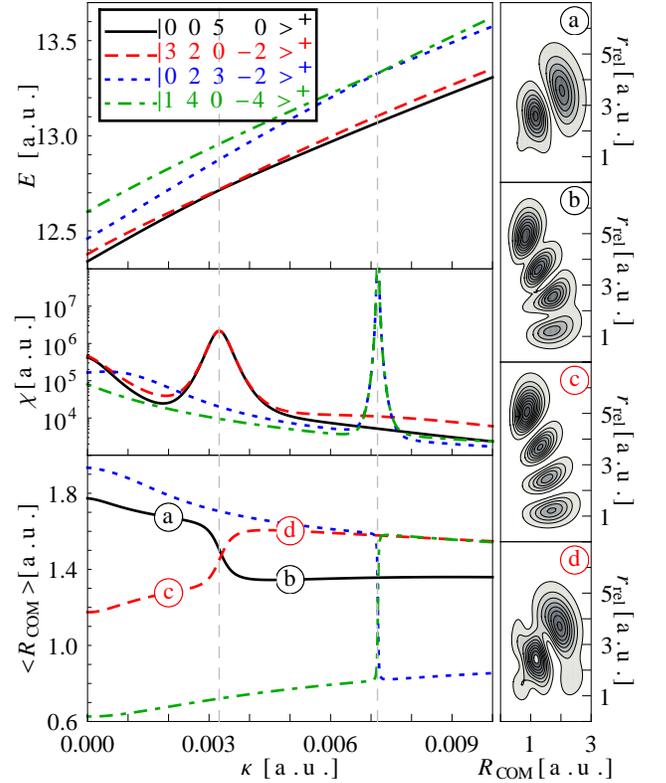}
 \caption{\label{fig_quantum} (Color online)  Upper panel: Example of two avoided crossings in the regime of weak quartic potential. The plot shows eigenenergies for four singlet states with even parity and vanishing total angular momentum in the potential with $\omega=1.0$, $\gamma=1.0$ and varying $\kappa$. The eigenstates are labeled by the exact quantum numbers  $n_c$, $m_c$, $n_r$, and $m_r$ of the harmonic problem ($\kappa=0$).
Middle panel: the quantum fidelity susceptibility for the same states. The susceptibilities of the two non-crossing states show peaks, which coincide at the maximum and are interchanged after the peak. 
Lower panel: The expectation values of the center of mass radial distance $\langle R\rangle$ in the vicinity of the avoided crossings of the upper panel.
Side panel: Contour plots of the states $|0,0,5,0\rangle^+$ and $|3,2,0,\text{-}2\rangle^+$ for $\kappa=0.002$ before the avoided crossing ((a) and (c), respectively), and for $\kappa=0.005$ after the avoided crossing ((b) and (d), respectively).}
\end{figure}

A manifestation of the non-separability of a system and its complexity is the occurrence of avoided crossings in the spectrum depending on a slowly varying parameter. This is the case in our system when $\kappa$ is the adiabatic parameter (see upper panel in Fig. \ref{fig_quantum}). Avoided crossings are naturally related to a drastic interchange of the symmetry properties of the eigenstates involved. This is typically characterized by abrupt changes in the behavior of some expectation values. For instance, as the eigenstates $|n_c, m_c, n_r,m_r\rangle^{\epsilon_p}=|1,4,0,\text{-}4\rangle^+$ and $|0,2,3,\text{-}2\rangle^+$ of the harmonic ($\kappa=0$) quantum dot {\em evolve} as $\kappa$ varies, there is a sudden jump in the expectation value of the COM radial distance $\langle R\rangle$ close to the avoided crossing around $\kappa=0.00716$ as shown in  the lower panel in Fig. \ref{fig_quantum}. Less pronounced is the change of $\langle R\rangle$ for the states $|0,0,5,0\rangle^+$ and $|3,2,0,\text{-}2\rangle^+$  close to the avoided crossing around $\kappa=0.00325$, though the respective wave functions \footnote{The wave functions are obtained with the help of the coordinate representation of the basis states as described in Appendix \ref{sec_coord_rep}.}  completely interchange their properties as seen in the side insets of Fig. \ref{fig_quantum}.

For the construction of the distribution $P_{\rm ac}(c)$ we require an efficient determination of avoided crossings in a large amount of spectral data. Though jumps in the expectation values $\langle R\rangle$ can be used to identify avoided crossings, the systematic detection of these jumps is not a trivial task. Alternatively, the {\em quantum fidelity susceptibility}  of the eigenstates provides an efficient method for such purpose. The quantum fidelity susceptibility  $\chi$ of an eigenstate $\psi_n$ is equivalent to the curvature of $\psi_n$ depending on the varying parameter \cite{Plo11}. It can be calculated 
via the static quantum fidelity $F_{\delta\kappa}(\kappa,n)=\left|\langle \psi_{n,\kappa} \mid\psi_{n,\kappa+\delta\kappa}\rangle\right|^2$,
\begin{align*}
   \chi&=\lim_{\delta\kappa\rightarrow0} \frac{1-F_{\delta\kappa}}{(\delta\kappa)^2}\approx-\lim_{\delta\kappa\rightarrow0}\frac{\log(F_{\delta\kappa})}{(\delta\kappa)^2},
\end{align*}
and is largely independent of the perturbation $\delta\kappa$ \cite{ploetzdiss}.

The typical behavior of the quantum fidelity susceptibility close to an avoided crossing is illustrated in the middle panel of Fig. \ref{fig_quantum}. It is characterized by three properties which are easy to implement for practical purposes: The susceptibility has a peak near an avoided crossing, this peak is nearly identical for the two non-crossing states and the mean values of the susceptibilities before and after the peak interchange for these two states. 

After checking that the susceptibility is independent of the perturbation for several values of $\delta\kappa\in\lbrace10^{-6},10^{-7},10^{-8},10^{-9}\rbrace$ data collection was performed using $\delta\kappa=10^{-7}$. With this method it was possible to determine nearly 9000 avoided crossings within a range of $0.0\le\kappa\le0.01$ and energies up to $70$. The levels start to mix depending on the quartic potential strength $\kappa$, though for small values of $\kappa$ the mixing can only be observed for high lying energy states.\\
For each avoided crossing the energy gap between the two close lying states, known as the width of the avoided crossing, is calculated and a statistical analysis of these values is performed. The expected distribution is the weighted sum of the distribution of the widths of the avoided crossings for the chaotic case with a $\delta$-peak, which represents the non-avoided crossings of the regular case \cite{Yan93}:
\begin{equation}\label{eq_pc}
 P(c)=(1-\lambda)\delta(c)+\frac{2\lambda^2}{\langle c\rangle\pi}\exp\left(\frac{-\lambda^2c^2}{\langle c\rangle^2\pi}\right),
\end{equation}
where $\langle c\rangle$ represents the mean value of the widths of the avoided crossings.  We fit with the cumulated distribution following Ref. \onlinecite{Yan93}.

\begin{figure}[t]
 \includegraphics[width=\columnwidth]{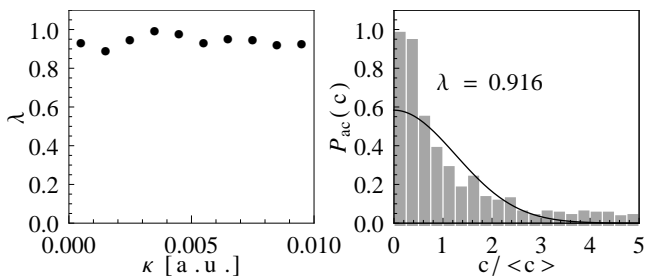}
 \caption{\label{fig_ac_dist}Fitting parameter $\lambda$ for the distribution of the widths of the avoided crossings for approximately the first 1000 eigenstates of the system. The right plot shows an example for the numerically determined distribution (grey bars) in the region $0.002\le \kappa\le0.003$ and the fit Eq. \eqref{eq_pc} without the $\delta$-peak close to zero (black line).}
\end{figure}

Even for small values of the quartic potential strength a high value for the fitting parameter $\lambda$ indicates a high degree of chaoticity in the system, see Fig. \ref{fig_ac_dist}. Again the very onset of the anharmonic term seems to carry the main effect, while the value of the interaction strength does not significantly change the distribution. The same argument as in the previous section holds here, as we analyze a wide energy regime capable to represent the features of anharmonicity even for small values of $\kappa$. The number of avoided crossings is already high for small values of $\kappa$, because the levels, originally degenerate in the purely harmonic case, strongly mix as soon as $\kappa\ne0$. 

\subsubsection{\label{sec_evl}Eigenvector localization}
We calculate an eigenstate specific information entropy depending on the parameters $\kappa$ and $\gamma$, which depends on the basis representation of the eigenvectors. Following the arguments by Zelevinsky \cite{Zel96} we obtain physically relevant results by choosing a reference basis for the representation of the eigenvectors that is physically related to the system under consideration. In our case this is the purely harmonic oscillator described in section \ref{sec_pho}. A general eigenvector of this system is given by
\begin{equation*}
|\psi_{nk}\rangle,\,n\in \{0,..,N\},\,k\in\{1,..,K_n\},
\end{equation*}
where $N$ goes to infinity and the values of $K_n$ are given in Table \ref{tab_deg_pxy}. We calculated numerically the harmonic case in order to achieve a complete ($K_n$) basis representation for all results presented here. Instead of determining the entropy directly in the harmonic oscillator basis, we project on the energy subspaces. For a general vector $|\varphi\rangle$, it holds
\begin{align*}
|\varphi\rangle&=\sum_{n=0}^{\infty} \sum_{k=1}^{K_n}c_{nk}|\psi_{nk}\rangle=\sum_{n=0}^{\infty}e_n|E_n\rangle,\end{align*}
with 
\begin{align*}
|E_n\rangle&=\frac{1}{e_n}\sum_{k=1}^{K_n}c_{nk}|\psi_{nk}\rangle,
\end{align*}
and the coefficients 
\begin{align*}
e_n&=\sqrt{\sum_{k=1}^{K_n}|c_{nk}|^2}
\end{align*}
chosen such that the representation basis $|E_n\rangle$ is orthonormal.
\begin{figure}[h]
 \includegraphics[width=\columnwidth]{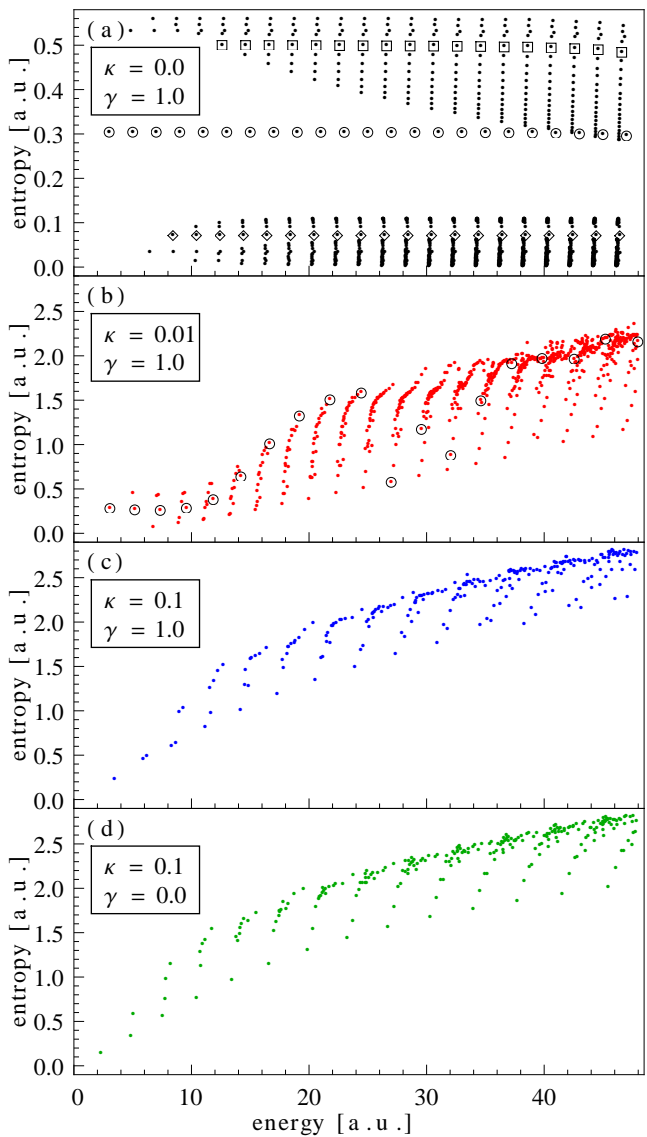}
 \caption{\label{fig_entropy}(Color online) Information entropies of states with singlet symmetry and even parity for four combinations of the parameters $\kappa$ and $\gamma$: (a) $\kappa=0.1$, $\gamma=0$, (b)  $\kappa=0.1$, $\gamma=1$, (c) $\kappa=0.01$, $\gamma=1$, and (d) $\kappa=0$, $\gamma=1$. In all cases there is a harmonic confinement with $\omega=1$. The entropy has been calculated for approximately 1200 eigenstates in (a), 700 eigenstates in (b) and 250 eigenstates in (c) and (d). In (a) we marked the states with the quantum numbers $|n_c,m_c,n_r,m_r\rangle^{\epsilon_p}=|k,0,0,0\rangle^+$  with circles (also in (b)), $|k,0,5,0\rangle^+$ with squares and $|k,2,1,\text{-}2\rangle^+$ with diamonds, where $k$ is an integer number. The other symmetry classes exhibit similar behavior.}
\end{figure}

The information entropy defined by 
\begin{align*}
S_\varphi&=\sum_{n=1}^{N}|e_n|^2\log(|e_n|^2),
\end{align*}
 is a measure of the localization of the vector $|\varphi\rangle$ in the harmonic oscillator basis. Large values of $S_\varphi$ imply a large spread of the state $|\varphi\rangle$ in this basis.

Fig. \ref{fig_entropy} shows the entropies of eigenstates with singlet symmetry and even parity of a two-electron quantum dot with a harmonic confinement ($\omega=1$) and four different situations. As our reference basis is the harmonic oscillator basis the purely harmonic case gave only vanishing entropies and is not considered here. For the harmonic case with electron-electron repulsion, in (a), a very regular behavior of the entropies can be observed. Indeed in this case the entropy is only a measure for the relative motion and the horizontally ordered entropies belong to states which differ only in the COM quantum numbers. The eigenstates contained in Table \ref{tab_harmonic} are highlighted with circles and systematically belong to the states within the region of higher entropy. These states are configurations with no angular momenta in the subsystems of the COM and of the relative motion. The second class of states, showing lower values for the entropy have non-vanishing angular momenta $m_c$ and 
$m_r=-m_c$ in the subsystems. We have marked a series of these states with diamonds.

In Fig. \ref{fig_entropy} (b) we show the entropies with weak anharmonic confinement ($\kappa=0.01$) and Coulomb repulsion. The behavior of the lowest eigenstates doesn't change significantly, up to an energy of $E\approx10$, while above this limit the entropies increase and lose most of their regular structure. The former effect can clearly be related to the narrowing of the potential and the higher number of harmonic states necessary to represent the eigenstates. The latter is a signature of the rising complexity of the system, induced by the coupling of the two previously separate motions and the increasing number of avoided crossings.  The occurrence of avoided crossings can in particular be seen from the jumps in the entropies of the states marked with circles, which we obtained by adiabatically following the levels from (a) to (b).

These effects are further enhanced for the stronger anharmonic ($\kappa=0.1$) case in (c), where only the ground state entropy is close to the harmonic case. 
This is in contrast to the case (d) with anharmonic confinement but without electron-electron interaction. The values of the entropies are as high as in case (c), but they still show a specific structure. This can be understood by considering that, in this case, the system is again separable in two independent particles.

Our analysis of the information entropy shows again, that the complexity stems from the interplay of the interparticle interaction and the anharmonic confinement. As long as one of those is omitted a high degree of regularity is still present, reflected in the behavior of the entropies. Qualitatively similar observations are obtained when we choose for instance the basis (\ref{basis0}) as a reference.

\section{\label{sec_con}Conclusion}
We have presented a detailed description of a numerically exact treatment of planar two-electron quantum dots, which extends the common harmonic model by introducing a quartic potential term. Our approach provides an accurate characterization of the spectrum of this system for a wide range of different parameters and symmetry classes. This has been exploited for studying anharmonic effects in the complexity of this system. Our analysis showed that generally the interplay between the Coulomb interaction and the anharmonic term is responsible for a significant reduction of the regions of regular classical motion of the planar quantum dot, which otherwise is integrable. On the quantum level, signatures of this mixed regular-chaotic underlying dynamics are observed in the level spacing distributions, the appearance of avoided crossings and the distribution of their separations, and in the eigenvector localization entropies. The complexity arises as soon as the anharmonicity is introduced in the 
interacting harmonic quantum dot. The complex quantum regime at rather high energies in the case of small anharmonic perturbations is shifted to lower energies as the anharmonicity increases. Apart from this, a common feature observed is the rather low dependence of the studied complexity measures on the strength $\kappa$ of the quartic confinement.

Our approach can be readily used for the study of further phenomena in two-electron quantum dots including the effects of the anharmonicity in Hund's rules and the consequences of chaos in the entanglement of two electrons. Our approach is also suitable for the study of decoherence processes in quantum dots. For instance, the anomalous behavior of quantum fidelity decay found in many-body systems \cite{Man08a,Man09} can be tested for a complex anharmonic system without assuming mean-field approximations.
Furthermore, the algebraic representation of observables, in particular the dipole operator, can be easily implemented. This offers the opportunity to investigate phenomena connected to the interactions of quantum dots with laser pulses.

\section*{Acknowledgments}
The authors are indebted to Sandro Wimberger, Bernard Piraux and Harald Friedrich for many interesting discussions.
They also thank traveling support provided by the COST Action CM0702 (CUSPFEL) of the European Union. S. Sch. is grateful for the support of the TUM graduate school.
\appendix
\section{\label{sec_units}Modified atomic units}
We apply modified atomic units ($\gamma\equiv1$), where we set $\hbar=m_*=\left(4\pi\epsilon_0\epsilon\right)^{-1}=q_e=k_B=1$. The natural scales are the modified Bohr-radius $$a_0=\frac{4\pi\epsilon_0\epsilon\hbar^2}{m_*q_e^2},$$ the modified Hartree energy $$E_{\rm h}=\frac{m_*q_e^4}{\left(4\pi\epsilon_0\epsilon\hbar\right)^2},$$ the natural frequency $$\nu_0=\frac{E_{\rm h}}{\hbar}=\frac{m_*q_e^4}{\hbar^3\left(4\pi\epsilon_0\epsilon\right)^2},$$ and the natural timescale $$ t_0=\frac{\hbar}E_{\rm h}=\frac{\hbar^3\left(4\pi\epsilon_0\epsilon\right)^2}{m_*q_e^4}.$$
Solid state quantum dots are most commonly realized on substrates of Indium Arsenide (InAs) and Gallium Arsenide (GaAs). For both cases we supply the values of the effective electron mass $m_*$, the dielectric constant $\epsilon$ and the natural scales:
\begin{itemize}
 \item InAs: $m_*\approx0.023\;m_e$, $\epsilon\approx15.15$, $a_0\approx35\,{\rm nm}$,\\
 $E_{\rm h}\approx2.7\,{\rm meV}$, $\nu_0\approx\,4.14\,{\rm THz}$, $t_0\approx0.24\,{\rm ps}$,
\item GaAs: $m_*\approx0.063\;m_e$, $\epsilon\approx12.9$, $a_0\approx11\, {\rm nm}$,\\
 $E_{\rm h}\approx10\,{\rm meV}$, $\nu_0\approx\,15.65\,{\rm THz}$, $t_0\approx0.066\,{\rm ps}$.
\end{itemize}
A typical value for the confining potential is \mbox{$E_0=\hbar\omega_0=3\,{\rm meV}$}, Ref. \onlinecite{Kou01}, which implies that $\omega$ is of order unity. The ratio of the harmonic confining potential to the natural energy scale of one Hartree ($E_{\rm h}$) is expressed as $\omega=E_0/E_{\rm h}$. We set $\omega \equiv 1$ in our calculations.

\begin{widetext}
\section{\label{sec_coord_rep}Coordinate basis representation}
In the polar coordinates associated to the parabolic coordinates \eqref{eq_ct} the basis functions can be represented in coordinate space with identification of the quantum numbers:
\begin{align*}
  M=n_1-n_2,& & L=n_3-n_4,  &&N=\min\left(n_1,n_2\right),&& K=\min\left(n_3,n_4\right),
\end{align*}
as follows:
\begin{equation*}
  \varphi_{\boldsymbol n}(r_+,\phi_+,r_-,\phi_-)=\langle n_1,n_2,n_3,n_4\mid r_+,\phi_+,r_-,\phi_-\rangle=\mathcal{N}\, r_+^{\left|M\right|}\,r_-^{\left|K\right|}\,L_N^{\left|M\right|}\left(r_+^2\right)\,L_K^{\left|L\right|}\left(r_-^2\right)\,e^{-\frac{1}{2}\left(r_+^2+r_-^2\right)}\,e^{-{\rm i}\left(M\phi_++L\phi_-\right)},
\end{equation*}
where $L_n^k(x)$ are associated Laguerre polynomials and $\mathcal{N}$ is a normalization constant:
\begin{equation*}
  \mathcal{N}=\left(-1\right)^{N+K}\frac{2}{\pi}\sqrt{\frac{N!\;K!}{\left(\left|M\right|+N\right)!\;\left(\left|L\right|+K\right)!}}.
\end{equation*}
The density plots shown in Fig. \ref{fig_quantum} are calculated by integrating over the angles in COM and relative coordinates, where the appropriate coordinate transformation is $r_+^2=4\,R_{\rm COM}$ and $r_-^2=2\,r_{\rm rel}$. For a general vector $|\psi\rangle=\sum_{\boldsymbol n} a_{\boldsymbol n} |{\boldsymbol n}\rangle$ we plot the following function:
\begin{equation*}
 \sum_{\boldsymbol n}\sum_{\boldsymbol m}a_{\boldsymbol n}^*a_{\boldsymbol m}\int_0^{\pi}d\phi_+\int_0^{\pi}d\phi_-\,R_{\rm COM}\,r_{\rm rel}\,\varphi_{\boldsymbol n}(\sqrt{4\,R_{\rm COM}},\phi_+,\sqrt{4\,r_{\rm rel}},\phi_-)^*\,\varphi_{\boldsymbol m}(\sqrt{4\,R_{\rm COM}},\phi_+,\sqrt{4\,r_{\rm rel}},\phi_-).
\end{equation*}

\end{widetext}


\end{document}